\pdfoutput=1
\documentclass[useAMS,usenatbib,fleqn]{mnras}
\usepackage{graphicx,subfig,lscape}
\usepackage[usenames,dvipsnames]{xcolor}
\usepackage{times,natbib}

\newcommand{\vs}{$v_{\rm e} \sin i$}
\newcommand{\teff}{$T_{\rm eff}$}
\newcommand{\lgg}{$\log\,{g}$}

\title[The surface structure of HD 32633]{The magnetic field topology and chemical abundance distributions of the Ap star HD 32633\thanks{
Based on observations obtained at the Canada-France-Hawaii Telescope (CFHT) which is operated by the National Research Council of Canada, 
the Institut National des Sciences de l'Univers of the Centre National de la Recherche Scientifique of France,  and the University of Hawaii.  Also based on observations obtained at the Bernard Lyot Telescope (TBL, Pic du Midi, France) of the Midi-Pyr\'en\'ees Observatory,  which is operated by the Institut National des Sciences de l'Univers of the Centre National de la Recherche Scientifique of France.}
}

\author[J. Silvester, O. Kochukhov and G.A. Wade]
{J. Silvester$^{1}$, O. Kochukhov$^{1}$ and G.A. Wade$^{2}$ \\
$^{1}$Department of Astronomy and Space Physics, Uppsala University, 751 20, Uppsala, Sweden\\
$^{2}$Department of Physics, Royal Military College of Canada, P.O. Box 17000, Station `Forces', Kingston, Ontario, Canada, K7K 7B4\\
}  
\begin{document}

\date{Accepted . Received }

\pagerange{\pageref{firstpage}--\pageref{lastpage}} \pubyear{2015}

\maketitle

\label{firstpage}

\begin{abstract}
Previous observations of the Ap star HD\,32633 indicated that its magnetic field was unusually complex in nature and could not be characterised by a simple dipolar structure. Here we derive magnetic field maps and chemical abundance distributions for this star using full Stokes vector (Stokes $IQUV$) high-resolution observations obtained with the ESPaDOnS and Narval spectropolarimeters. Our maps, produced using the {\sc Invers10} magnetic Doppler imaging (MDI) code, show that HD\,32633 has a strong magnetic field which features two large regions of opposite polarity but deviates significantly from a pure dipole field. We use a spherical harmonic expansion to characterise the magnetic field and find that the harmonic energy is predominately in the $\ell=1$ and $\ell=2$ poloidal modes with a small toroidal component. At the same time, we demonstrate that the observed Stokes parameter profiles of HD\,32633 cannot be fully described by either a dipolar or dipolar plus quadrupolar field geometry. We compare the magnetic field topology of HD\,32633 with other early-type stars for which MDI analyses have been performed, supporting a trend of increasing field complexity with stellar mass. We then compare the magnetic field topology of HD\,32633 with derived chemical abundance maps for the elements Mg, Si, Ti, Cr, Fe, Ni and Nd. We find that the iron-peak elements show similar distributions, but we are unable to find a clear correlation between the location of local chemical enhancements or depletions and the magnetic field structure. 
\end{abstract}


\section{Introduction}
The magnetic chemically peculiar (Ap/Bp) intermediate-mass stars are main sequence objects of spectral types A and B exhibiting a number of unusual properties when compared to normal stars of similar spectral types. Their defining characteristic is the presence of strong, globally organised surface magnetic fields with a strength ranging from a few hundred G to a few tens of kG. In the first approximation these fields have a geometry of a dipole inclined with respect to the stellar rotational axis \citep{Donati09}. The presence of such strong magnetic fields leads to a grossly non-solar atmospheric chemistry and to the formation of high-contrast horizontal (spots) and vertical (stratification) chemical abundance inhomogeneities. As a consequence of this non-uniform surface structure, Ap/Bp stars exhibit a profound spectroscopic and photometric rotational variability, even though the intrinsic configuration of spots and magnetic field remains constant on the time scale of decades.

Detailed empirical information on the geometry of magnetic and chemical surface structures in Ap/Bp stars is essential for understanding the physical mechanisms governing evolution of the global magnetic fields in stellar interiors \citep{Braithwaite06,Duez10} and for testing theories of chemical segregation by radiative diffusion \citep{Leblanc09,Alecian10}. Aiming to collect this information for a meaningful sample of Ap/Bp stars, we are reconstructing magnetic and chemical maps of their surfaces from high-resolution spectropolarimetric observations \citep{Kochukhov04,Kochukhov11,Kochukhov14,Kochukhov15,Kochukhov10,Silvester14a,Rusomarov15}, with a special emphasis on the interpretation of line profile variability in all four Stokes parameters. In the this paper we present the first such analysis for the Ap star HD\,32633.

The star HD\,32633 (HZ\,Aur, HIP\,23733) has an effective temperature of 12,000--13,000~K and is of the peculiar subclass SiCr \citep{Renson09}. Early observations by \citet{Babcock58} measured the magnetic field as a function of phase; this author noted that the magnetic field was strong, reversing in sign approximately once a week and that the field behaved erratically at the negative polarity. Later observations by \citet{Borra80} and \citet{Leone00} found that the longitudinal field variation as a function of rotational phase could not be fit with a simple sinusoidal curve. Instead, the curve appeared non-harmonic, with extrema separated by 0.6 of the rotational period. These early observations indicated that the magnetic field of HD\,32633 departs from a simple dipolar structure. It was later suggested by \citet{Glagolevskij08} that the magnetic field of HD\,32633 could be modeled assuming two dipoles at opposite regions of the star, located near the rotational equator. HD 32633 does not show signs of pulsational variability in either the roAp \citep{Joshi06} or SPB \citep{Aerts14} frequency ranges.

These observations establish HD\,32633 as an interesting target for mapping using magnetic Doppler imaging (MDI), from which we can obtain detailed information on the magnetic field geometry and thus better understand the unusual non-sinusoidal longitudinal field phase curve behaviour of this star. Furthermore, the pioneering MuSiCoS four Stokes parameter observations of Ap stars by \citet{Wade00a} demonstrated that HD\,32633 shows the strongest linear polarisation signatures among the Ap/Bp stars included in their survey. More recently, \citet{Silvester12} collected a superb new data set of full Stokes vector observations of several Ap/Bp stars, including HD\,32633, using the ESPaDOnS and Narval instruments.

Spectropolarimetric data in four Stokes parameters can be successfully interpreted with the magnetic Doppler imaging method, in particular with the {\sc Invers10} code \citep[see][]{Piskunov02,Kochukhov02} optimised for simultaneous reconstruction of the vector magnetic maps and chemical abundance distributions. A remarkable result of the application of MDI to four Stokes parameter spectra of Ap stars was the discovery that the magnetic topologies of at least some Ap stars depart significantly from low-order multipolar configurations and contain important small-scale components \citep{Kochukhov04,Kochukhov10,Silvester14a}. However, after $\alpha^2$\,CVn (HD\,112413) studied by \citet{Silvester14a}, HD\,32633 is only the second Ap star to be mapped using the modern high-resolution Stokes $IQUV$ data of \citet{Silvester12} and as such provides an exciting further insight into how the magnetic field topologies of Ap stars might deviate from the canonical dipolar geometry. In addition to investigating the magnetic field topology we also take advantage of the time-resolved spectropolarimetric data to map the chemical abundance structures on the surface of HD\,32633 in order to examine correlations between the magnetic field topology and the locations of any abundance enhancements or depletions. 

The paper is organised as follows. Section~\ref{sect:obs} briefly describes the observational data. Section~\ref{sect:params} discusses derivation of stellar parameters. Section~\ref{sect:methods} summarises the procedure of magnetic Doppler imaging. The resulting magnetic and chemical abundance maps are presented in Sections~\ref{sect:mag} and \ref{sect:abn}, respectively, and are discussed in the context of other MDI studies and theoretical investigations in Section~\ref{sect:discuss}.

\begin{figure*}
\begin{center}
 \includegraphics[height=\textwidth, angle=90]{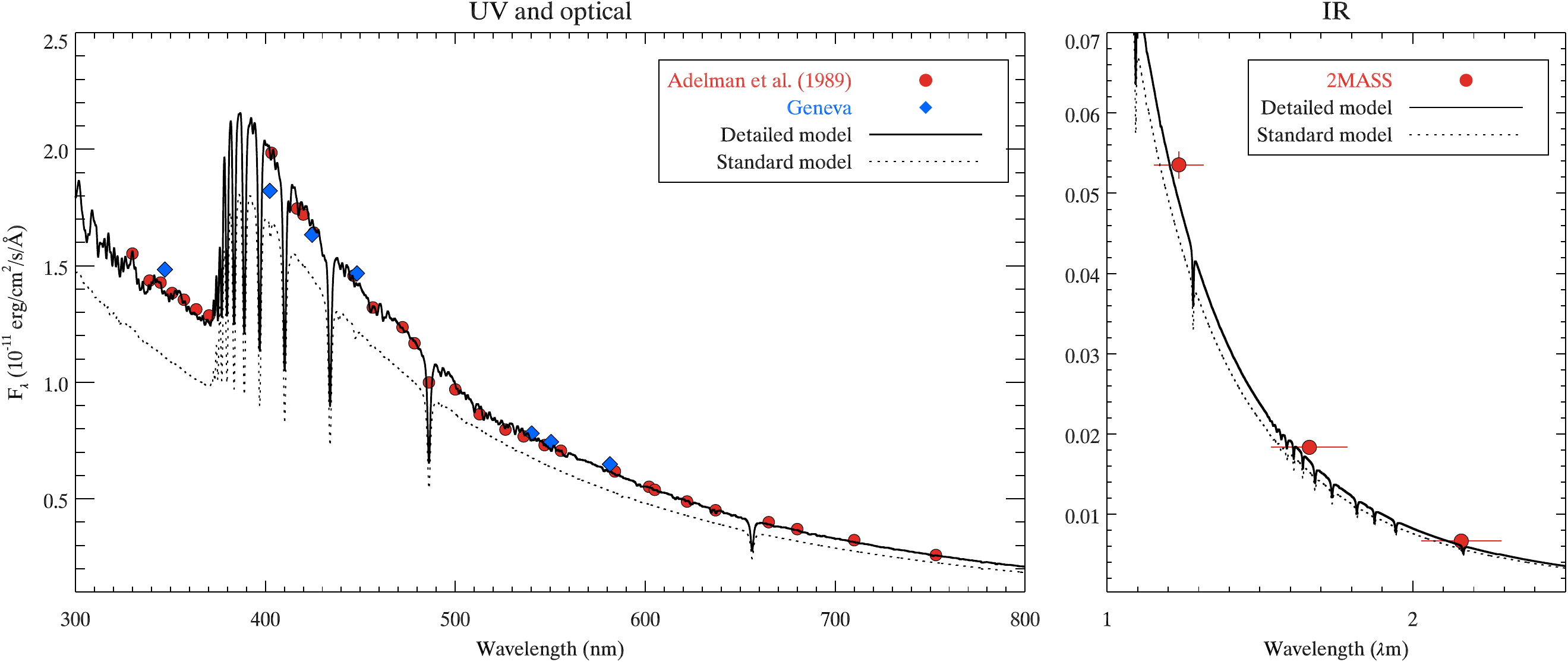}
  \caption{Comparison between theoretical spectral energy distributions (lines) and observations (symbols) in the UV, optical and near-IR spectral regions. The two theoretical SEDs correspond to the calculations for $T_{\rm eff}=12,800$, $\log g=4.15$ including effects of individual non-solar abundances of HD\,32633 and magnetic field (solid line) and standard model (solar abundances, no magnetic field) for the same $T_{\rm eff}$ and $\log g$ (dashed line).}
\label{SED}
\end{center}
\end{figure*}

\begin{figure}
\begin{center}
 \includegraphics[width=0.47\textwidth]{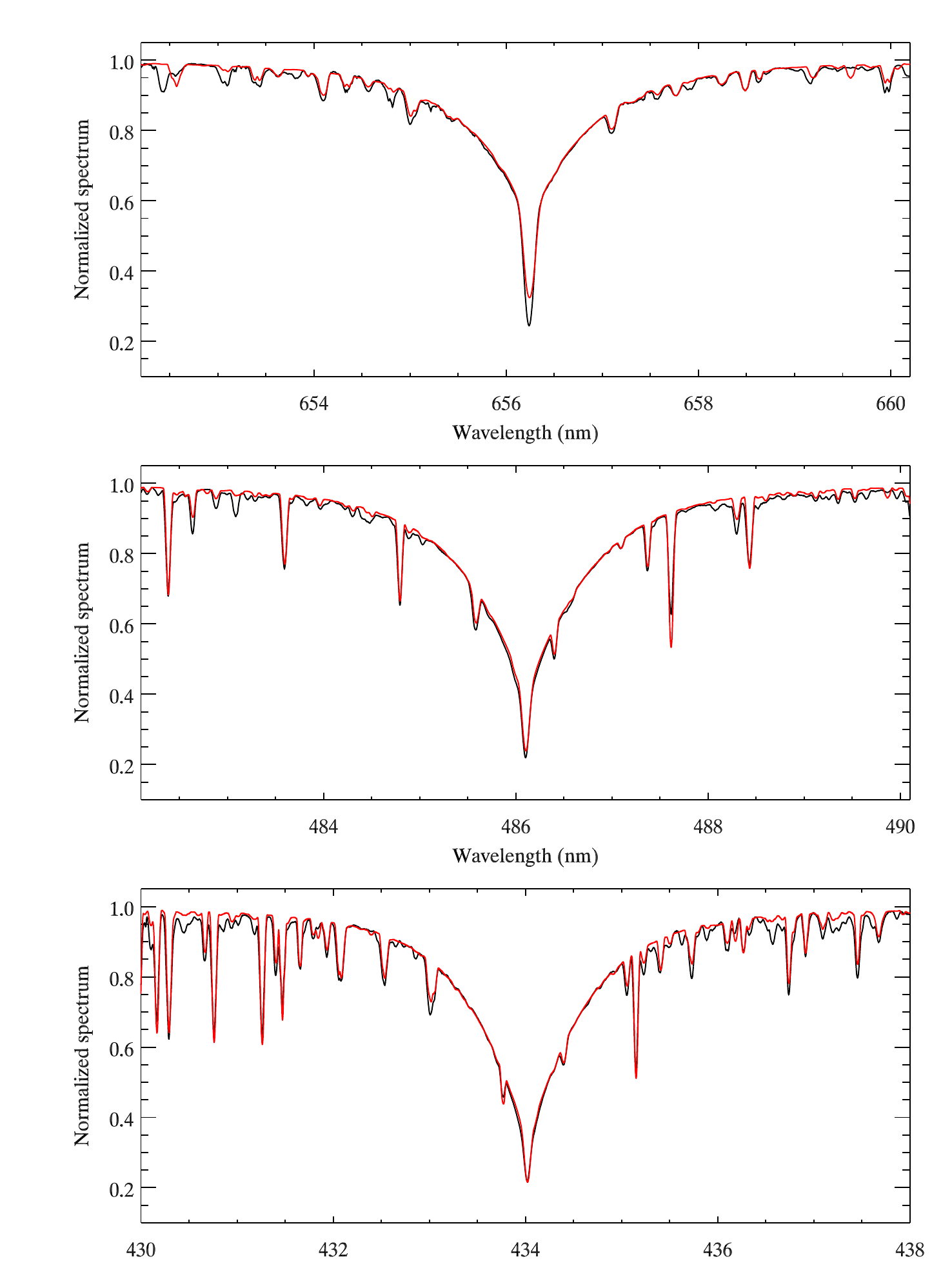}
  \caption{Comparison between the observed (dark curve) and computed (bright curve) profiles of the H$\alpha$, H$\beta$, and H$\gamma$ lines in the spectrum of HD\,32633.}
\label{Hlines}
\end{center}
\end{figure}

\section{Spectropolarimetric observations}
\label{sect:obs}

Observations of HD\,32633 were obtained between 2006 and 2010, with both ESPaDOnS and Narval spectropolarimeters in all four Stokes parameters (Stokes $IQUV$). The full details of the observations, including the observing log, are reported by \citet{Silvester12}. Briefly, the spectra have a resolving power of 65,000, cover the wavelength range 369--1048~nm, and have a typical signal-to-noise ratio of 600:1. A total of 20 full Stokes $IQUV$ observations are available for HD\,32633, with an additional Stokes $IV$ observation. Observations sample on average every 0.1 of a phase,  with closer sampling at certain phases.  As described by \citet{Silvester12}, the reduction of the Stokes $IQUV$ spectra was performed on-site at the respective observatories using the {\sc Libre-Esprit} reduction package \citep{Donati97}. Normalisation of the spectra was performed order by order using an {\sc IDL} code optimised for Ap stars. The rotational phases were computed according to the ephemeris of \citet{Adelman97}.

\section{Fundamental stellar parameters}
\label{sect:params}

We obtained an initial estimate of the atmospheric parameters of HD\,32633 by applying the calibrations by \citet{Netopil08} to the $uvby\beta$ \citep{Hauck98} and Geneva \citep{Rufener76} photometric observations of this star. Both sets of photometric observations indicate $T_{\rm eff}=12500$--13000~K and $\log g=4.0$--4.3. Guided by these estimates, we computed a model atmosphere structure for $T_{\rm eff}=13000$~K, $\log g=4.0$ and $[M/H]=+1.0$~dex using the {\sc LLmodels} code \citep{Shulyak04}. Using this model atmosphere we then determined abundances for about a dozen chemical species, including He, Mg, Si, Fe-peak and several rare-earth elements, by fitting the {\sc Synmast} \citep{Kochukhov10b} theoretical spectra to the phase-averaged Stokes $I$ spectrum of HD\,32633. The resulting individual chemical abundances were adopted for calculation of a new {\sc LLmodels} atmospheric grid covering the relevant $T_{\rm eff}$ and $\log g$ ranges. In these calculations we took into account modification of the line opacity due to Zeeman splitting and polarised radiative transfer \citep{Khan06} assuming the field modulus of 10~kG \citep{Glagolevskij08}.

The final effective temperature was determined by fitting the spectral energy distribution (SED) in the optical (Geneva photometry converted to absolute fluxes and spectrophotometry by \citet{Adelman89}) and near-IR (2MASS fluxes) wavelength regions, using the {\sc LLmodels} code to model the SED. HD\,32633 is known to be somewhat reddened, with $E(B-V)$ ranging from 0.05 to 0.12 according to different estimates the literature {\citep[e.g][]{Adelman80}}. Here we adopted $E(B-V)=0.10\pm0.02$ as a compromise value. Figure~\ref{SED} illustrates comparison of the observed SED with the model fluxes corresponding to $T_{\rm eff}=12800$~K and $\log g=4.15$. The latter value of the surface gravity was established by comparing {\sc Synmast} calculations to the observed profiles of hydrogen Balmer lines. As evident from Fig.~\ref{Hlines}, our choice of atmospheric parameters yields an excellent fit to the H$\alpha$, H$\beta$, and H$\gamma$ lines in the spectrum of HD\,32633. 


As a by-product of the SED fitting we determined an angular diameter and converted it to the stellar radius, $R=1.99\pm0.25R_\odot$, using the parallax $\pi=5.57\pm0.67$~mas \citep{vanLeeuwen07}. Our radius value agrees well with previous estimates by \citet[][$1.92\pm0.32R_\odot$]{Kochukhov06} and \citet[][$2.4R_\odot$]{Leone00}.

A summary of fundamental stellar parameters determined or adopted in our study is given in Table~\ref{tbl:params}.

\begin{table}
\begin{center}
\caption{Fundamental parameters of HD\,32633.\label{tbl:params}}
\begin{tabular}{ccc}
\hline
\hline  
Parameter & Value & Reference \\
\hline
\teff &  $12800 \pm 500$ K  &  this study \\
\lgg & $4.15 \pm 0.1$ &  this study \\ 
$R$ & $1.99 \pm 0.25$ $R_\odot$ &  this study \\ 
\vs & $18.0 \pm 1.0$ km\,s$^{-1}$ & this study \\
$i$ & $80\degr \pm 5\degr$ & this study \\
$\Theta$ & $90\degr \pm 5\degr$ & this study \\
$P_{\rm rot}$ & 6.43000 d & \citet{Adelman97} \\
\hline
\label{parameter-table}
\end{tabular}
\end{center}
\end{table}


\begin{table}
\begin{center}
\caption{Atomic lines used for the mapping of HD 32633. The $\log gf$ values are those provided by the Vienna Atomic Line Database \citep[VALD][]{Kupka99}.
\label{tbl:lines}}

\begin{tabular}{lcc}
\hline
\hline
Ion & Wavelength & $\log gf$ \\
&(\AA )  & \\
\hline
\multicolumn{3}{c}{Magnetic and abundance mapping}\\
Si\,{\sc ii}& 6347.109 & 0.170 \\
Cr\,{\sc ii} & 4558.650 & -0.449  \\
                  & 4558.650 & -4.950  \\ 
         & 4588.199  & -0.627 \\
         & 4824.127 & -0.970   \\
Fe\,{\sc ii}& 4583.829 & -1.860 \\
             & 4923.921 & -1.320 \\
         & 5018.436 & -1.220  \\
         & 5169.028 & -1.250  \\
Nd\,{\sc iii}& 5050.695 & -1.060  \\
                  & 5102.428 & -0.620  \\
\hline
\multicolumn{3}{c}{Abundance mapping}\\
Mg\,{\sc ii}& 4481.126 & 0.740 \\
                  & 4481.150 & -0.560  \\   
                  & 4481.325 & 0.590  \\   
Si\,{\sc ii} & 4130.872 & -0.824 \\   
                  & 4130.894 & 0.563  \\                   
                    & 5055.984 & 0.530  \\   
  Ti\,{\sc ii}& 4290.215 & -0.870  \\   
                & 4563.757 & -0.690  \\   
 Ni\,{\sc ii}& 4067.031 & -1.834   \\     
                  \hline
\label{line-list}
\end{tabular}
\end{center}
\end{table} 

\begin{figure*}
\begin{center}
    \includegraphics[width=0.72\textwidth, angle=-90]{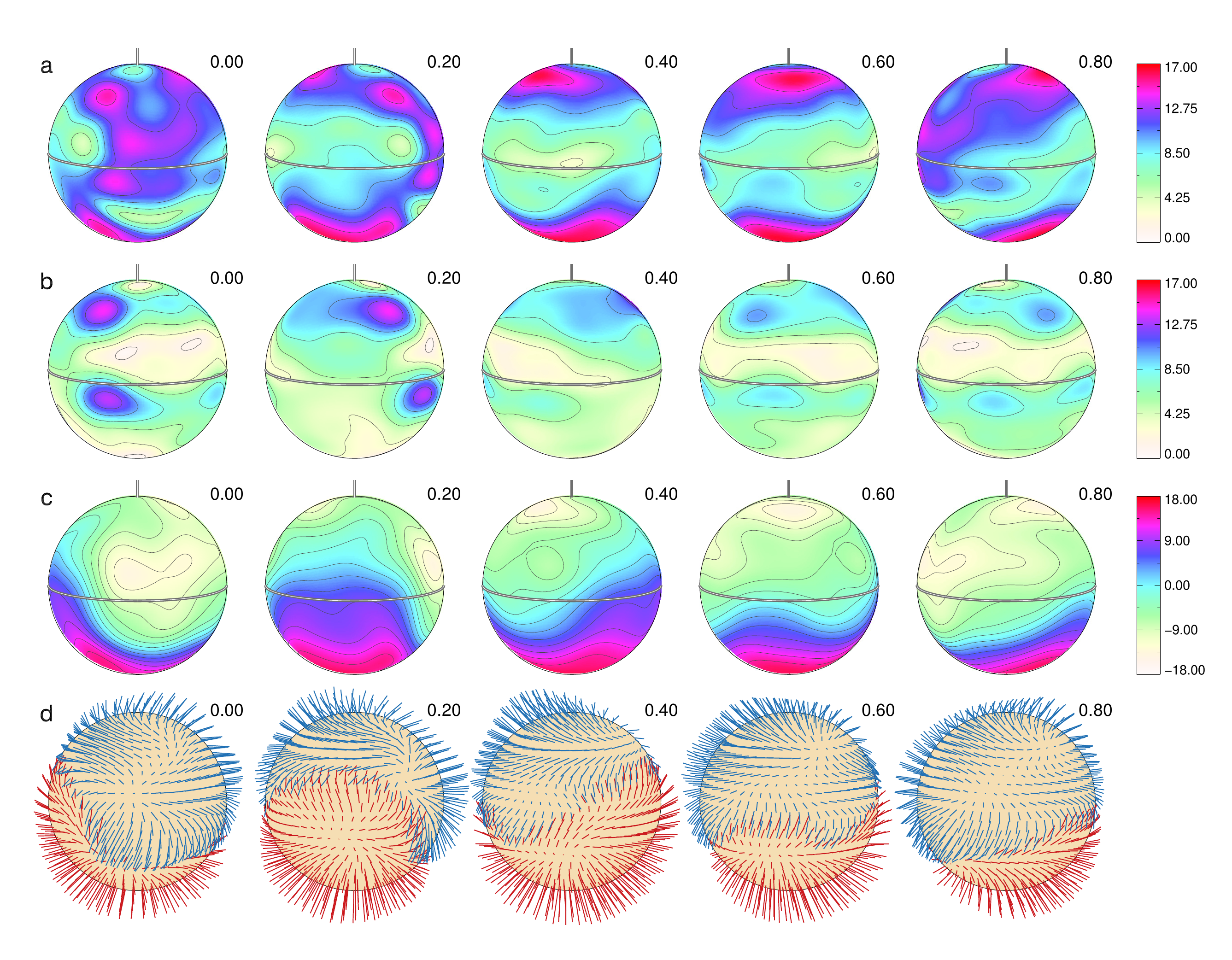}
  \caption{Surface magnetic field distribution of HD 32633 derived using Stokes $IQUV$ computed with {\sc Invers10} using a spherical harmonic description of the field topology. The spherical plots (at inclination angle of $ i = 80\degr$), show distributions of: a) the field modulus, b) the horizontal field, c) the radial field and d) the field orientation. Each column corresponds to a different phase of rotation (0.0, 0.2, 0.4, 0.6 and 0.8) and the colour bars on the right side indicate the magnetic field strength in kG.}
\label{Maps-field}
\end{center}
\end{figure*}

\section{Magnetic and chemical inversions}
\label{sect:methods}

An application of four Stokes parameter MDI requires knowledge of several other quantities in addition to the fundamental stellar parameters determined above. In particular, we need to known the projected rotational velocity \vs\ and to specify the inclination (tilt with respect to the line of sight) and azimuth (orientation in the plane of the sky) angles of the stellar rotational axis.  A review of the literature did not reveal a definitive value for the inclination angle of HD\,32633. Therefore to constrain the orientation of the rotational axis we ran a grid of inversions by varying both the inclination and azimuth angles. We determined a suitable value for both angles by finding the pair of angles which resulted in the smallest deviation between the model spectra and the observational data. Similarly, to determine an optimal \vs, we started with the previously determined value of $19.0 \pm 2.0$ km\,s$^{-1}$ \citep{Silvester12}, performed inversions for a range of \vs\ values and selected the value which provided the smallest deviation between the observations and model profiles. 

Through this process we found that the following parameters resulted in the best fit between the model and observations:  inclination $i=80\degr \pm 5\degr$, azimuth angle of $90\degr \pm 5\degr$ and \vs\,=\,$18.0 \pm 1.0 $ km\,s$^{-1}$. We note that our inclination angle agrees with the value of this parameter estimated by \citet{Borra80} ($77\degr$) and by \citet{Glagolevskij08} ($84\degr \pm 5\degr$). The rotational period from \citet{Adelman97} together with the stellar radius and \vs\ determined in our paper imply an inclination angle close to $90\degr$.

The methodology of magnetic Doppler imaging employed in this study is similar to that used in the mapping of the Ap $\alpha^2$\,CVn \citep{Silvester14a,Silvester14b}. We rely on the {\sc Invers10} code \citep{Piskunov02} to obtain best-fitting magnetic and chemical distributions. A regularisation function is used to facilitate convergence to a global chi-square minimum and guarantee that we find the simplest surface distribution that fits the data. A new feature of the magnetic field reconstruction is that we parameterise the magnetic field topology in terms of a general spherical harmonic expansion (up to the angular degree $\ell=10$) as described by \citet{Kochukhov14}. However, for comparison with our previous studies, we also performed inversions with discrete pixel maps for each of the three magnetic field components. In both cases the choice of regularisation parameter is essential. For the maps presented in this paper, the value of regularisation was chosen so that it gave the lowest total discrepancy between observations and the model and yet still reproduced the Stokes profiles without fitting to a significant amount of noise. An additional constraint was chosen, such that the total regularisation must be at the minimum only a factor of 10 times smaller than the total discrepancy. After this limit the improvement to the discrepancy becomes increasingly smaller, but the map will continue to become increasingly patchy.
For most of the inversions, a stepwise approach was taken with the code using five steps of regularisation (decreasing in value by a factor of three) during the course of one inversion run. The result is five sets of fits between observations and the model and five resulting maps (each corresponding to a different regularisation parameter). This allowed us to determine the most suitable value of regularisation to use for a given inversion run.   

Many metal lines in the ESPaDOnS and Narval spectra of HD\,32633 exhibit polarisation signatures suitable for modelling in four Stokes parameters. Previously we found that simultaneous interpretation of line profile variability of several chemical elements with diverse surface distributions yields the most robust magnetic maps \citep{Silvester14b}. Taking these results into account, we selected 10 mostly unblended lines of Si\,{\sc ii}, Cr\,{\sc ii}, Fe\,{\sc ii} and Nd\,{\sc iii} for the simultaneous magnetic and abundance mapping in Stokes $IQUV$ parameters. In addition, 6 lines of Mg\,{\sc ii}, Si\,{\sc ii}, Ti\,{\sc ii} and Ni\,{\sc ii} with comparatively weak Stokes $QU$ signatures were used for reconstruction of the surface abundance distribution of these elements from the Stokes $IV$ spectra adopting a fixed magnetic field geometry. The atomic parameters of 16 spectral features employed in the MDI of HD\,32633 are given in Table~\ref{tbl:lines}. For each spectral interval we also included relevant minor blends, which are omitted in this table.

\section{Magnetic Field Structure}
\label{sect:mag}

\begin{figure*}
\begin{center}
  \includegraphics[width=0.99\textwidth]{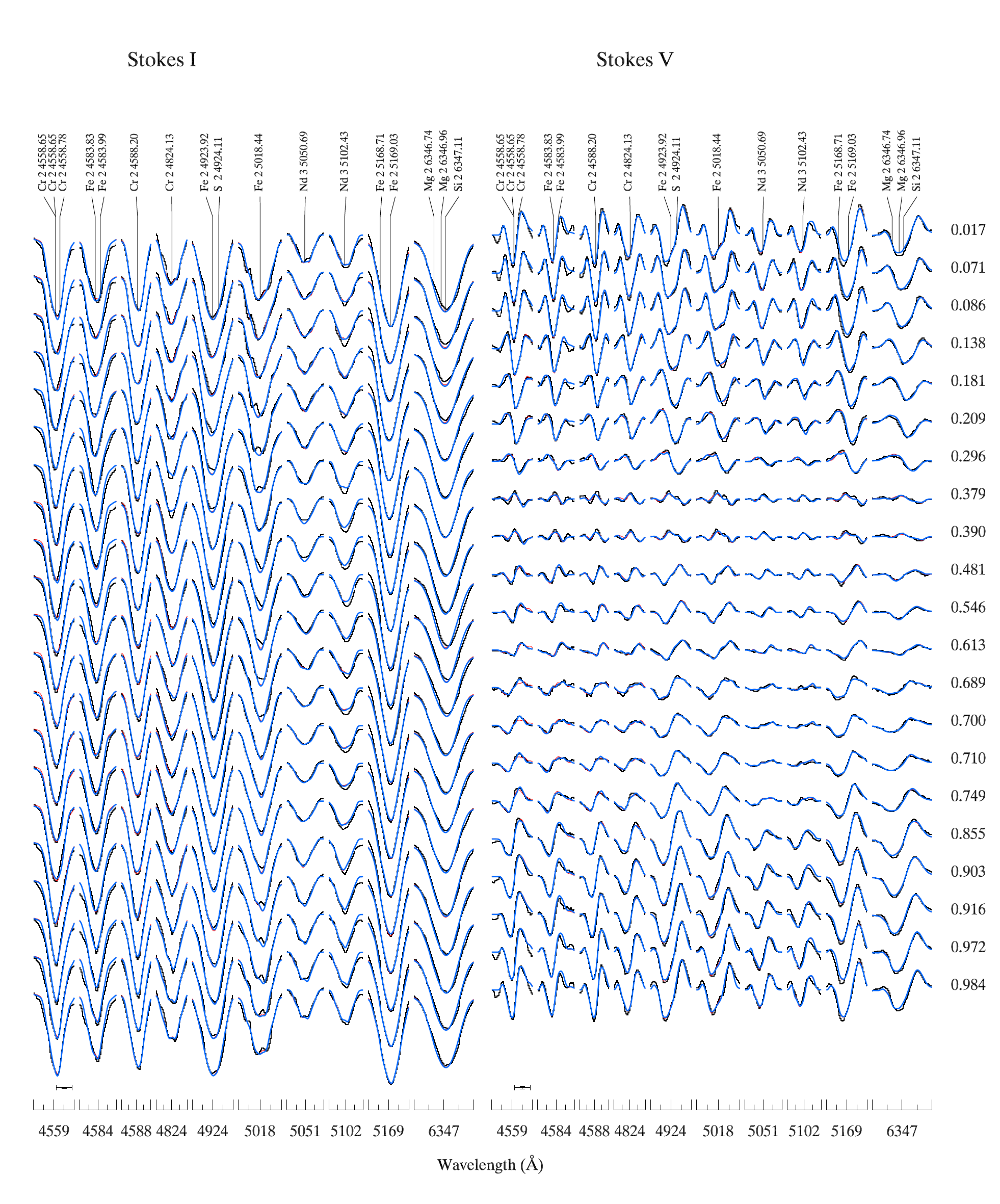}
   \caption{Stokes $IV$ profiles for all lines used in the magnetic field mapping. Observed spectra (black histogram line) are compared to the synthetic profiles corresponding to the spherical harmonic inversion (thick blue lines) and to the discrete surface element inversion (thin red line). The spectra are offset vertically according to the rotational phase indicated to the right of the Stokes $V$ panel. The bars in the lower left corners indicate the vertical and horizontal scales (1 \% of the Stokes $I$ continuum intensity and 0.5 \AA).}
\label{Fit-IV-Fld}
\end{center}
\end{figure*}

\begin{figure*}
\begin{center}
    \includegraphics[width=0.99\textwidth]{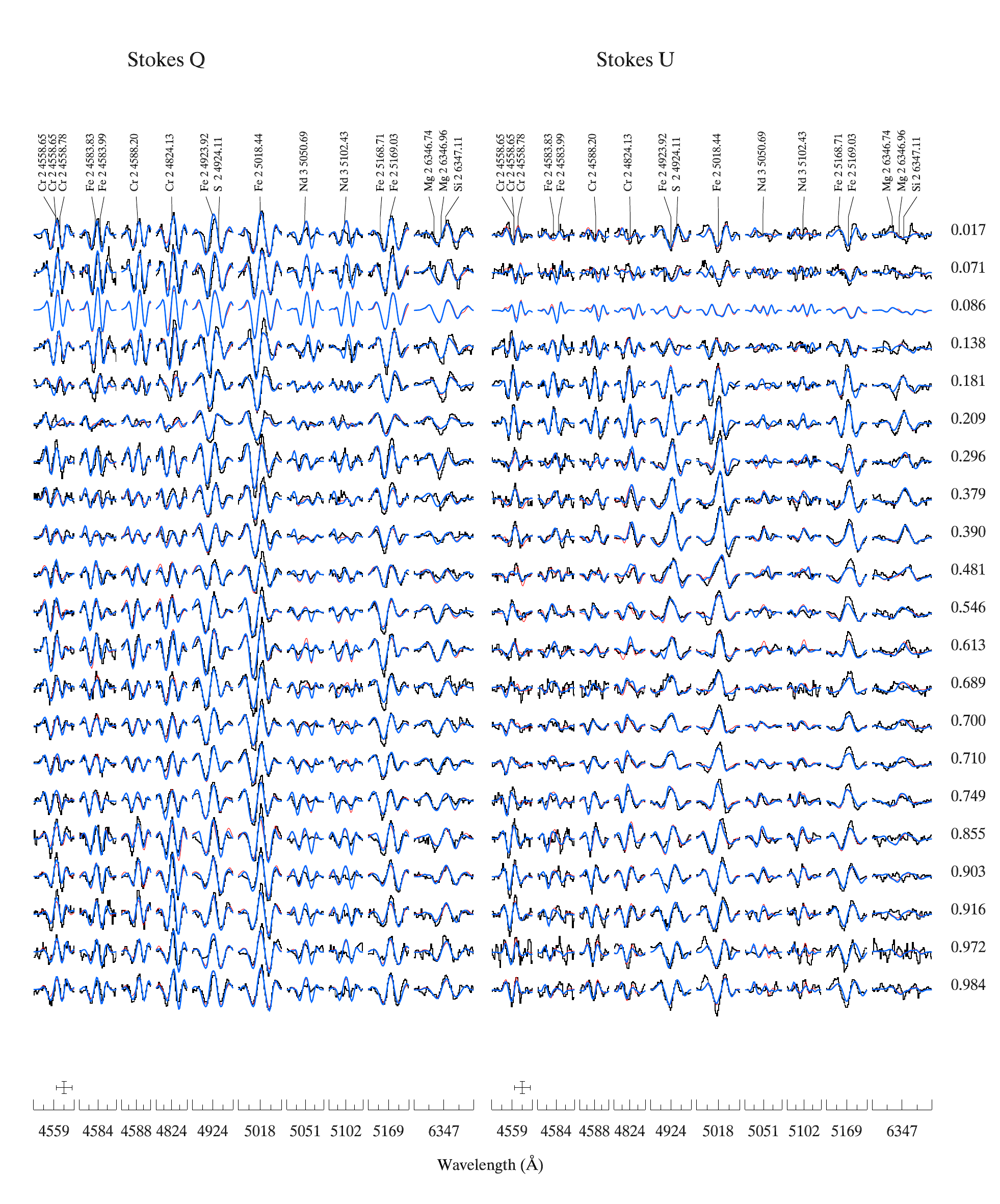}
   \caption{Same as Fig.~\ref{Fit-IV-Fld} for the Stokes $QU$ profiles.}
\label{Fit-QU-Fld}
\end{center}
\end{figure*}

To characterise the magnetic field topology of HD 32633, maps were derived using the spherical harmonic expansion version of {\sc Invers10} \citep{Kochukhov14} and, for an addition test, using the discrete surface element version of the code \citep{Piskunov02}. The final map discussed below was obtained using the spherical harmonic parameterisation. In this inversion we obtained the best fit between the observed Stokes $IQUV$ parameters and model profiles by allowing the code to fit harmonics up to $\ell=10$, including a toroidal component and allowing independent radial and horizontal poloidal field components. In other words, the three sets of spherical harmonic coefficients $\alpha_{\ell,m}$, $\beta_{\ell,m}$ and $\gamma_{\ell,m}$ specifying the surface field geometry \citep[see][]{Kochukhov14} were allowed to vary independently. The resulting magnetic field structure is presented in Fig.~\ref{Maps-field} and the corresponding fits to observations are illustrated in Figs.~\ref{Fit-IV-Fld} and \ref{Fit-QU-Fld}.

Our MDI modelling indicates that the field structure of HD\,32633 is dominated by the two large regions of opposite field polarity (Fig.~\ref{Maps-field}d), i.e. the field is approximately dipolar. However, this structure is noticeably distorted, in particular if one examines the distributions of the field modulus and horizontal field (Fig.~\ref{Maps-field}ab). Figure~\ref{Maps-field}a suggests that the surface field modulus ranges between about 4 and 17 kG. The corresponding disk-integrated mean field modulus varies between 6.8 and 7.7 kG depending on the rotational phase (somewhat less than was estimated by  \citet{Glagolevskij08}).

\begin{figure*}
\begin{center}
    \includegraphics[width=0.95\textwidth]{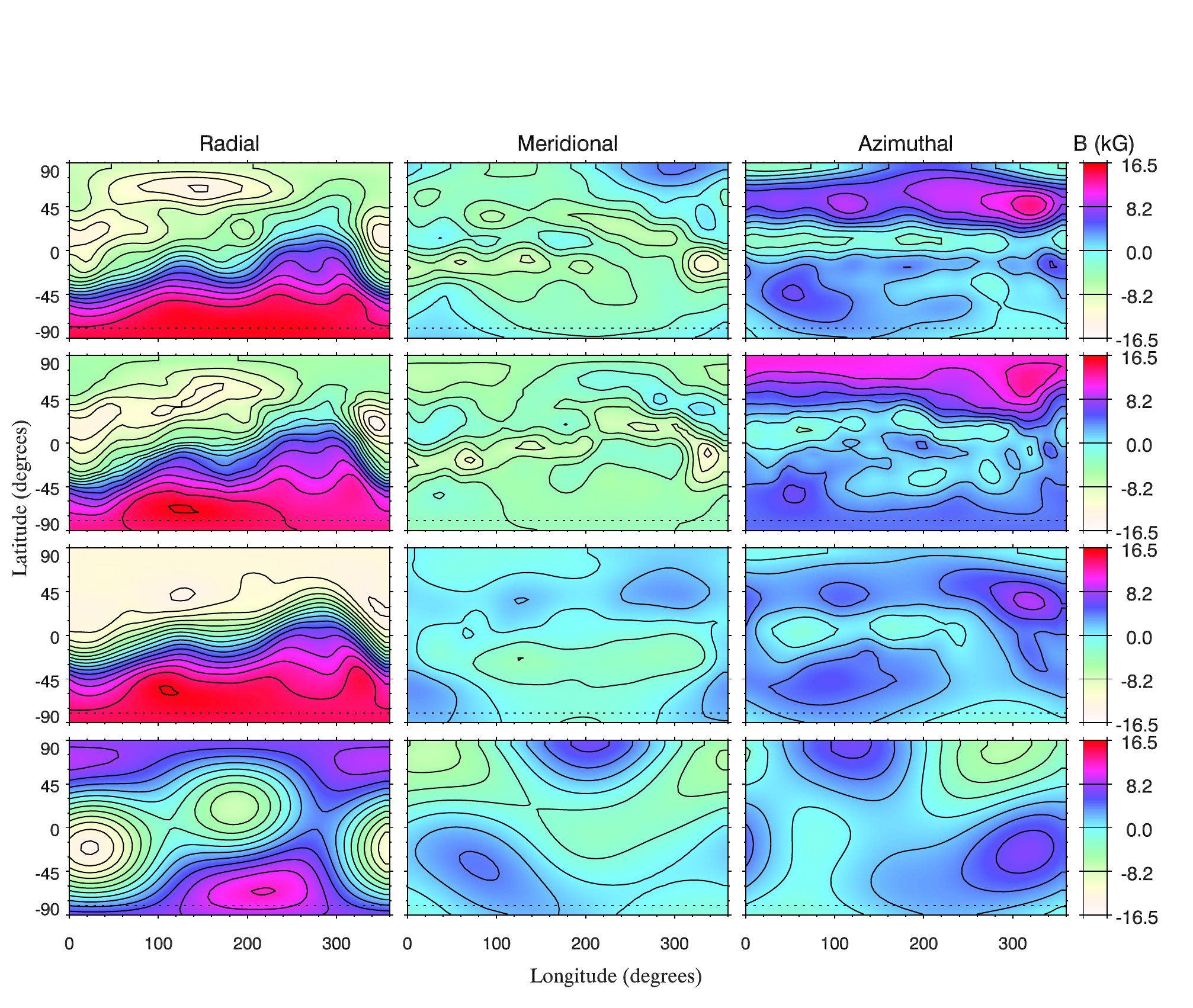}
      \caption{Comparison of the radial, meridional and azimuthal magnetic components for different MDI inversions. From top to bottom: map derived with the spherical harmonic inversion and using Stokes $IQUV$, map derived with the discrete surface element inversion and using Stokes $IQUV$, map derived with the spherical harmonic inversion and using only Stokes $IV$, map derived assuming a dipole + quadrupole structure and using Stokes $IQUV$. The dashed line indicates the lowest visible latitude for the adopted inclination angle $i = 80\degr$. A contour stepping of 2 kG has been used for a range of $[-16, +16]$~kG.}
\label{field-rec}
\end{center}
\end{figure*}

We also derived a magnetic field map using the inversion based on a discrete surface element description of the magnetic field. For this inversion to converge to a consistent solution it required us to initialise the inversions with a basic dipolar field structure. 
The resulting fit to the observed Stokes profiles is also shown in Figs.~\ref{Fit-IV-Fld} and \ref{Fit-QU-Fld}. Generally, there is little difference for the Stokes $IV$ profiles between these fits and those obtained using the spherical harmonic parameterisation. However, the spherical harmonic inversion does a marginally better job at reproducing the Stokes $QU$ profiles. A detailed comparison of the radial, meridional and azimuthal components of the discrete and spherical harmonic magnetic field maps is presented in Fig.~\ref{field-rec}. It is evident that the inferred field topology does not depend significantly on the parameterisation of the surface field map.

For an additional test, we derived a magnetic map using only Stokes $IV$ profiles. The purpose of this exercise was to assess how much of the field complexity is lost in the inversion ignoring linear polarisation data. The resulting magnetic field map is shown in Fig.~\ref{field-rec}, where it can be directly compared to the results of inversions based on all four Stokes parameter spectra. This comparison shows that the Stokes $IV$ map does not show as much detail compared to the Stokes $IQUV$ maps. This loss of complexity is similar to the results of previous MDI studies where the Stokes $IQUV$ and $IV$ inversion results were compared \citep{Kochukhov10,Rosen15}. We also find that the meridional field component is systematically stronger in the $IQUV$ maps. This is an expected result as seen from previous numerical tests \citep{Kochukhov02, Rosen12}.

\begin{figure*}
\begin{center}
  \includegraphics[width=0.99\textwidth]{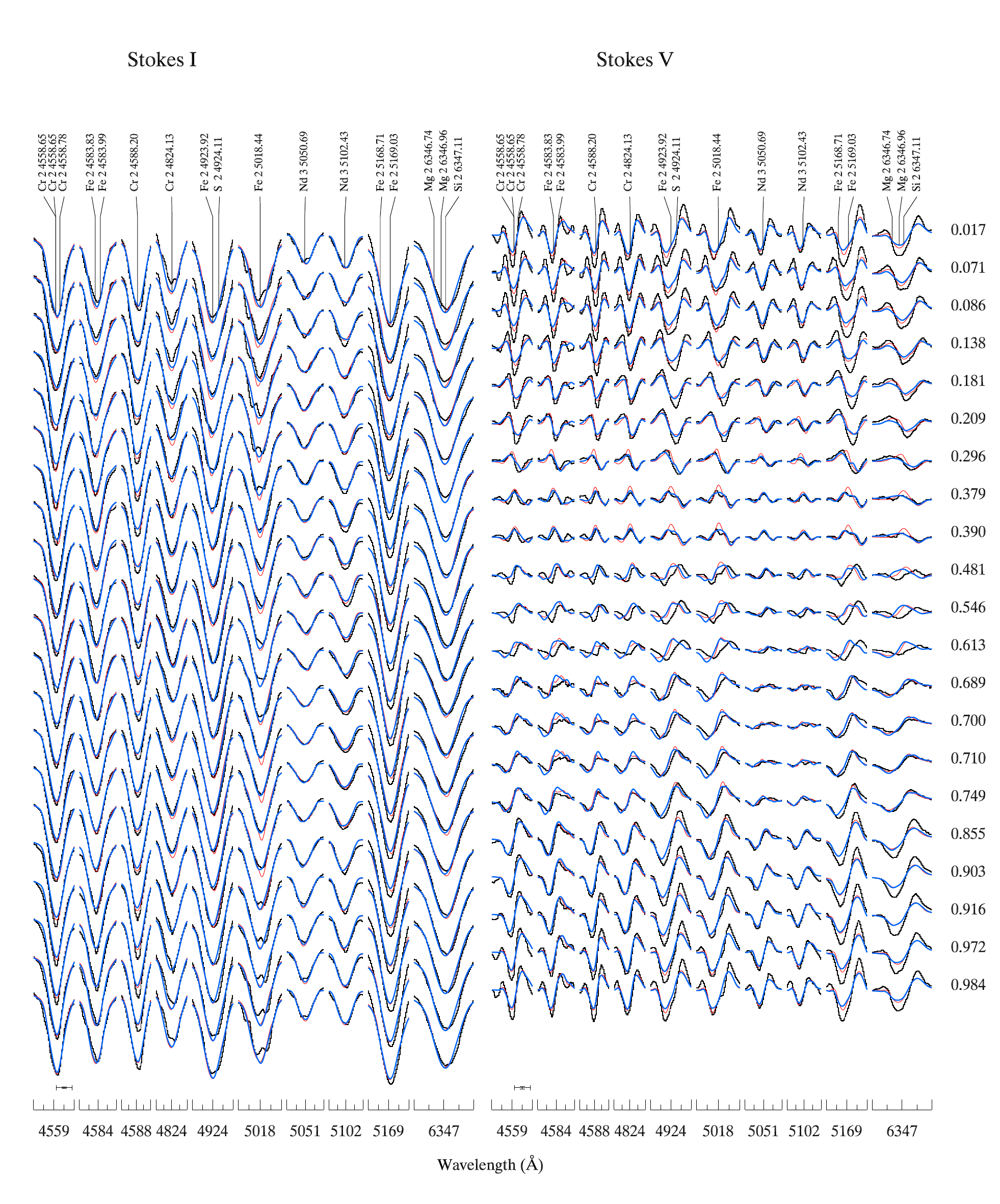}
   \caption{Same as Fig.~\ref{Fit-IV-Fld} for the comparison between observations (black histogram line), best fit dipolar model profiles (thick blue line) and best fit dipole + quadrupole model profiles (thin red line).}
\label{Fit-IV-Sp}
\end{center}
\end{figure*}

\begin{figure*}
\begin{center}
    \includegraphics[width=0.99\textwidth]{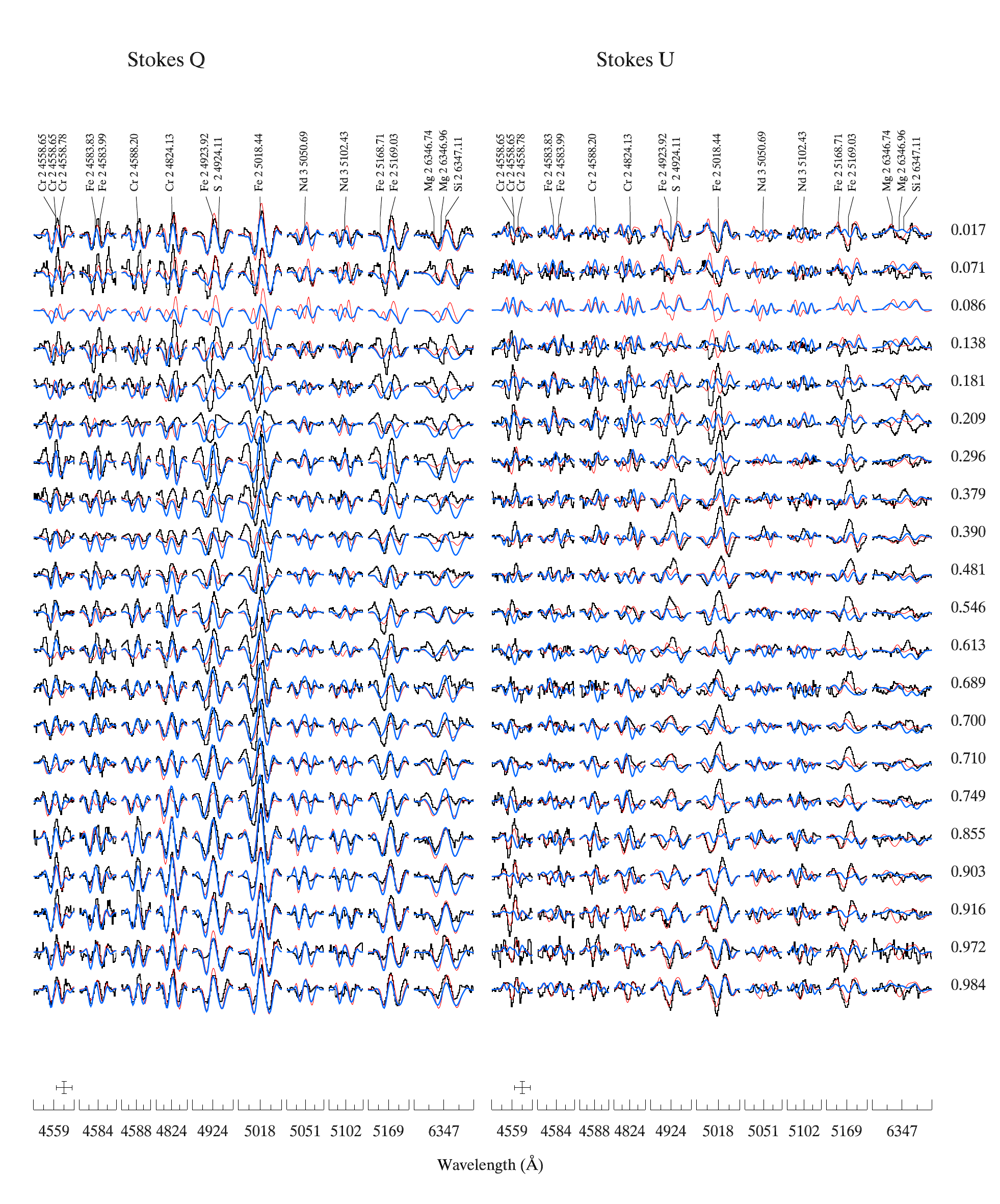}
   \caption{Same as Fig.~\ref{Fit-IV-Sp} for the Stokes $QU$ profiles.}
\label{Fit-QU-Sp}
\end{center}
\end{figure*}

To confirm the reality of smaller scale ($\ell \ge 3$) magnetic features in the MDI magnetic map of HD\,32633 we investigated if the observed Stokes profiles of this star could be reproduced by limiting the code to reconstruct a purely dipole ($\ell_{\rm max}=1$) or dipole + quadrupole ($\ell_{\rm max}=2$) field topologies. The best-fitting dipole + quadrupole field distribution is illustrated in Fig.~\ref{field-rec}. The corresponding fits to observations are shown in Figs.~\ref{Fit-IV-Sp} and \ref{Fit-QU-Sp}. These fits are clearly inferior when compared to those achieved by the inversions which allowed for a more complex field geometry. This result demonstrates that the magnetic field of HD\,32633 indeed deviates significantly from a pure dipole or dipole + quadrupole structure and that such low-order multipole models cannot reproduce the observed polarisation spectra. In particular, the Stokes $Q$ and $U$ profiles cannot be fit at all.

\begin{figure}
\begin{center}
    \includegraphics[width=0.50\textwidth]{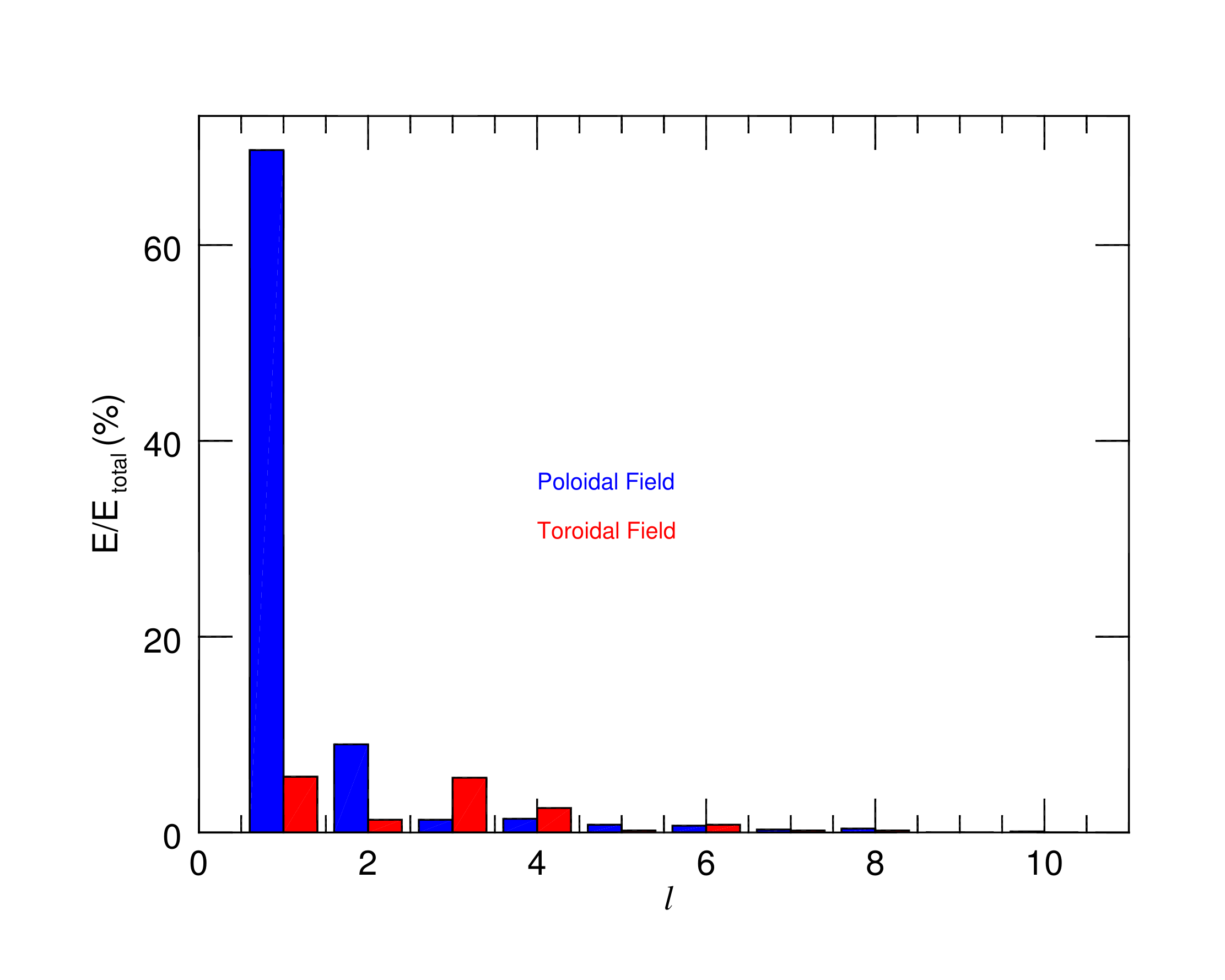}
   \caption{Distribution of the magnetic energy between different spherical harmonic modes in the MDI magnetic field map of HD\,32633. Bars of different colour represent the poloidal (blue/dark) and toroidal (red/light) magnetic components.}
\label{power-plot}
\end{center}
\end{figure}

The use of the spherical harmonic description of the magnetic field enables a detailed characterisation of the field structure at different spatial scales by examining relative magnetic energies of individual harmonic modes. The magnetic energy distribution for the final magnetic map is illustrated in Fig.~\ref{power-plot} as a function of the angular degree $\ell$. The relative harmonic energies also reported in Table.~\ref{energies}.  We find that the total contribution of the poloidal and toroidal field is 84\% and 16\% respectively. The dipole mode ($\ell=1$) contains 75\% of the magnetic energy; 10\% of the energy is in $\ell=2$ and the remaining 15\% is in $\ell \ge 3$.

\begin{table}
\begin{center}
\caption{Distribution of poloidal ($E_{\rm pol}$) and toroidal ($E_{\rm tor}$) magnetic field energy over different harmonic modes for the best-fitting MDI map of HD\,32633.}
\begin{tabular}{lccc}
\hline
\hline
$\ell$ & $E_{\rm pol}$ & $E_{\rm tor}$ & $E_{\rm tot}$ \\
 & (\%) & (\%) & (\%) \\
\hline
  1   & 69.7 & 5.7 &  75.4 \\
  2  & 9.0  &  1.3 &  10.3 \\
  3   & 1.3   & 5.6    & 6.9 \\
  4   &  1.4 &  2.5  & 3.9 \\
  5   & 0.8   & 0.2   & 1.0 \\
  6    & 0.7   & 0.8   & 1.5 \\
  7    & 0.3   &  0.2   & 0.4 \\
  8    & 0.4  & 0.2   & 0.6 \\
  9    & 0.0   & 0.0  & 0.1 \\
 10    & 0.1   & 0.0   & 0.1 \\
\hline
 Total &  83.5 &  16.5  & 100 \\
\hline
\label{energies}
\end{tabular}
\end{center}
\end{table} 

\begin{figure*}
\begin{center}
 \vspace{5 mm}
 \includegraphics[width=0.85\textwidth]{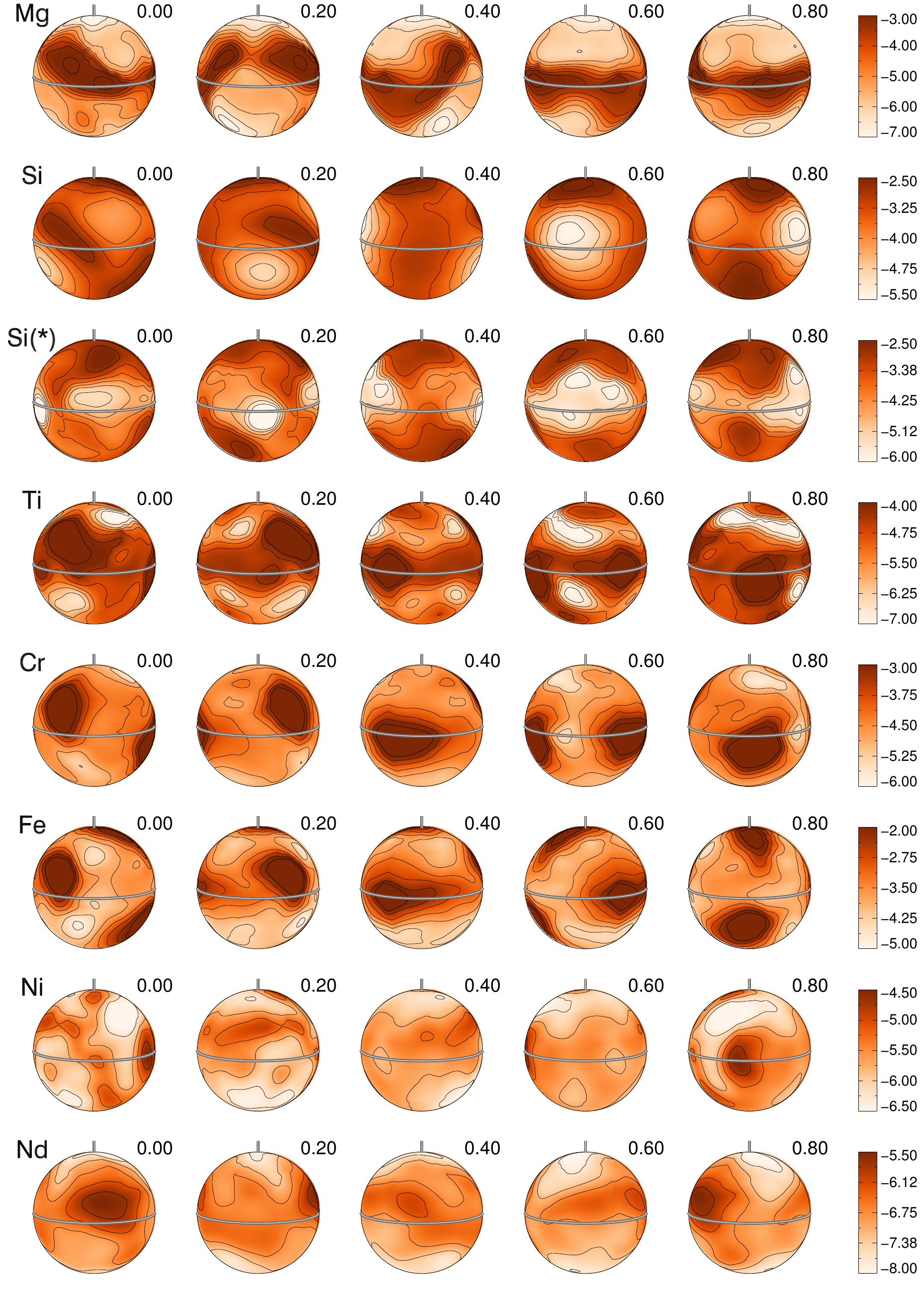}
    \caption{Chemical abundance distributions derived for singly ionised Mg, Si, Ti, Cr, Fe, Ni and doubly ionised Nd. Each column corresponds to a different rotational phase (0.0, 0.2, 0.4, 0.6 and 0.8). The solid line shows the location of the stellar equator. The visible rotational pole is indicated by the short thick line. Two silicon maps presented: Si from one line used in the Stokes $IQUV$ magnetic mapping and Si(*) from two lines in Stokes $IV$ abundance inversion. The colour bars on the right side indicate chemical abundance in $\log N_{\rm el}/\log N_{\rm tot}$ units. }
\label{Maps-Abn}
\end{center}
\end{figure*}

\section{Chemical Abundance Distributions}
\label{sect:abn}

The selection process for determining which chemical elements to map was very similar to that used by \citet{Silvester14b} for $\alpha^2$~CVn. However, unlike in the case of that star, the Stokes $I$ profiles of HD\,32633 exhibit relatively little variability as a function of rotational phase. Therefore lines were selected primarily on the basis that they did not show significant blending with other atomic species. 

The abundance distributions derived here were based on one of the following two inversions: abundance maps reconstructed simultaneously with the magnetic field geometry (based on spherical harmonic parameterisation of the field) and the Stokes $IQUV$ data (Si, Cr, Fe and Nd) and abundance maps based only on the Stokes $IV$ observations (Mg, Si, Ti and Ni) produced assuming the magnetic field structure as derived first from the Stokes $IQUV$ inversion. 

When visualising the final abundance maps in spherical plots, the scale was chosen such that extreme outliers were excluded. These outliers, typically comprising 2--12\% of all surface elements, were identified via the use of a histogram of abundance values for the entire stellar surface, following the procedure described by \citet{Silvester14b}.

\subsection{Magnesium and Silicon}

The magnesium map was produced using the Stokes $IV$ profiles of the Mg\,{\sc ii}\ $\lambda$ 4481 feature.  The fit between observations and the model can be seen in Fig.~\ref{Mg-Si-Ti-Ni-fit} and the resulting map is presented in Fig.~\ref{Maps-Abn}. The inferred magnesium surface abundance ranges from $-3.0$ dex to $-7.0$ dex  (where the solar value is $-4.44$) on the $\log (N_{\rm Mg}/N_{\rm tot})$ scale. The high abundance region is located in a large belt-like structure above the stellar equator at phases 0.0 to 0.4 and below the equator at other phases. The high abundance belt appears to trace areas where the radial magnetic field is generally weak.

For silicon we derived two maps. One was produced using the Stokes $IQUV$ profiles of the Si\,{\sc ii}\ $\lambda$ 6347 line and another was obtained from the Stokes $IV$ profiles of the Si\,{\sc ii} lines $\lambda$ 4130 and 5056. The comparison of observations and the model profiles can be seen in Fig.~\ref{Fit-IV-Fld}, Fig.~\ref{Fit-QU-Fld} and Fig.~\ref{Mg-Si-Ti-Ni-fit}. The resulting Si maps are shown in Fig.~\ref{Maps-Abn}. Examining the distribution of silicon in the two maps, the abundance values vary from $-2.5$ dex  to $-5.5$ dex and from $-2.5$ to $-6.0$ dex respectively (where the solar value is $-4.53$). The high abundance areas are located close to the rotational poles, with a large relative underabundance area centred on the stellar equator and clearly seen at phase 0.60 in both maps. 

Reconstruction of the silicon abundance distribution from the Stokes $IV$ spectra was somewhat complicated by the fact, while the $\lambda$ 5056 line could be fit using the VALD oscillator strength, the the $\lambda$ 4130 line was under-fit and required the oscillator strength to be increased by 0.06 dex.  The same problem was encountered by \citet{Kochukhov14} when modelling the same Si\,{\sc ii} lines in the spectrum of Ap star CU~Vir. 

Comparing the chemical map with the magnetic field distribution, in both silicon abundance maps the location of depleted abundance traces areas with weaker magnetic field. On the other hand, enhanced abundance areas correspond to regions close to where the magnetic field is strongest.  At least with regard to the location of areas of depleted silicon abundance,  this behaviour is similar to what was found for $\alpha^2$ CVn  \citep{Silvester14b}.

\subsection{Titanium, Chromium and Iron}

\begin{figure*}
\begin{center}
 \includegraphics[width=0.25\textwidth]{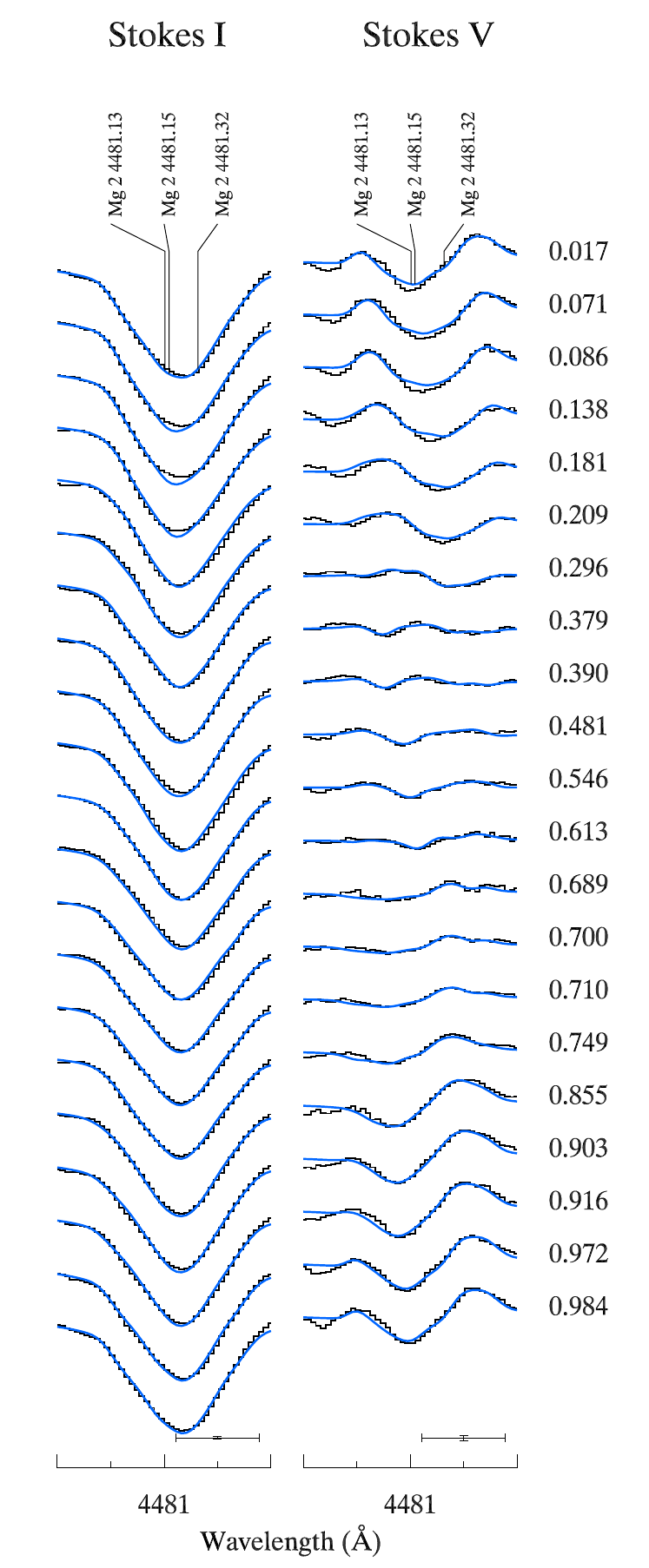}
  \includegraphics[width=0.24\textwidth]{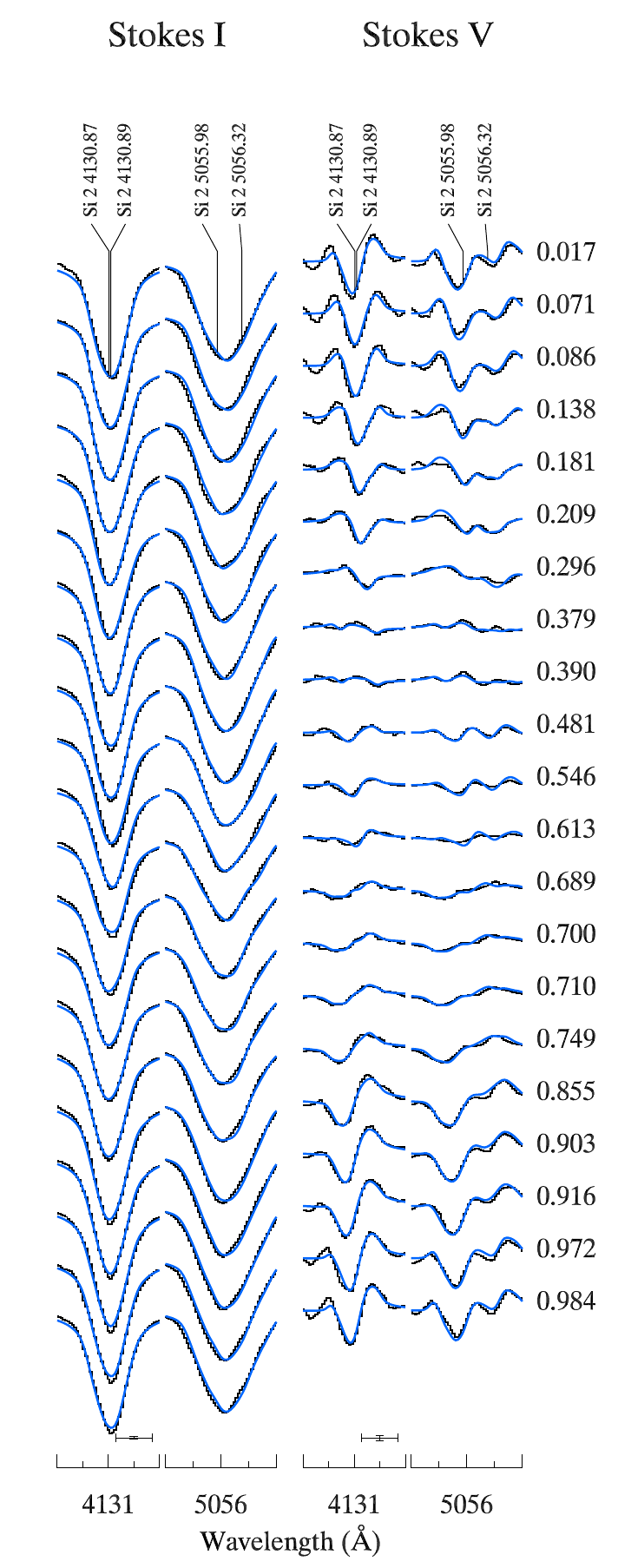}
   \includegraphics[width=0.24\textwidth]{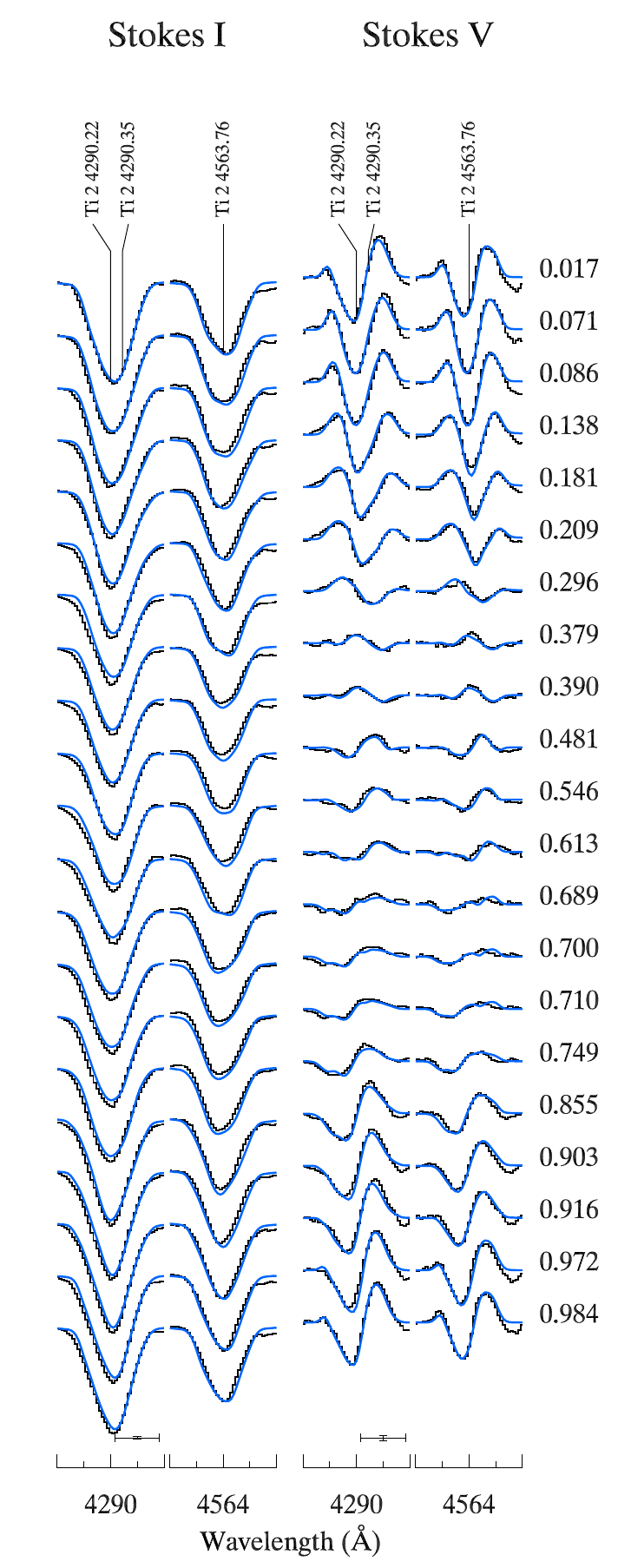} 
  \includegraphics[width=0.24\textwidth]{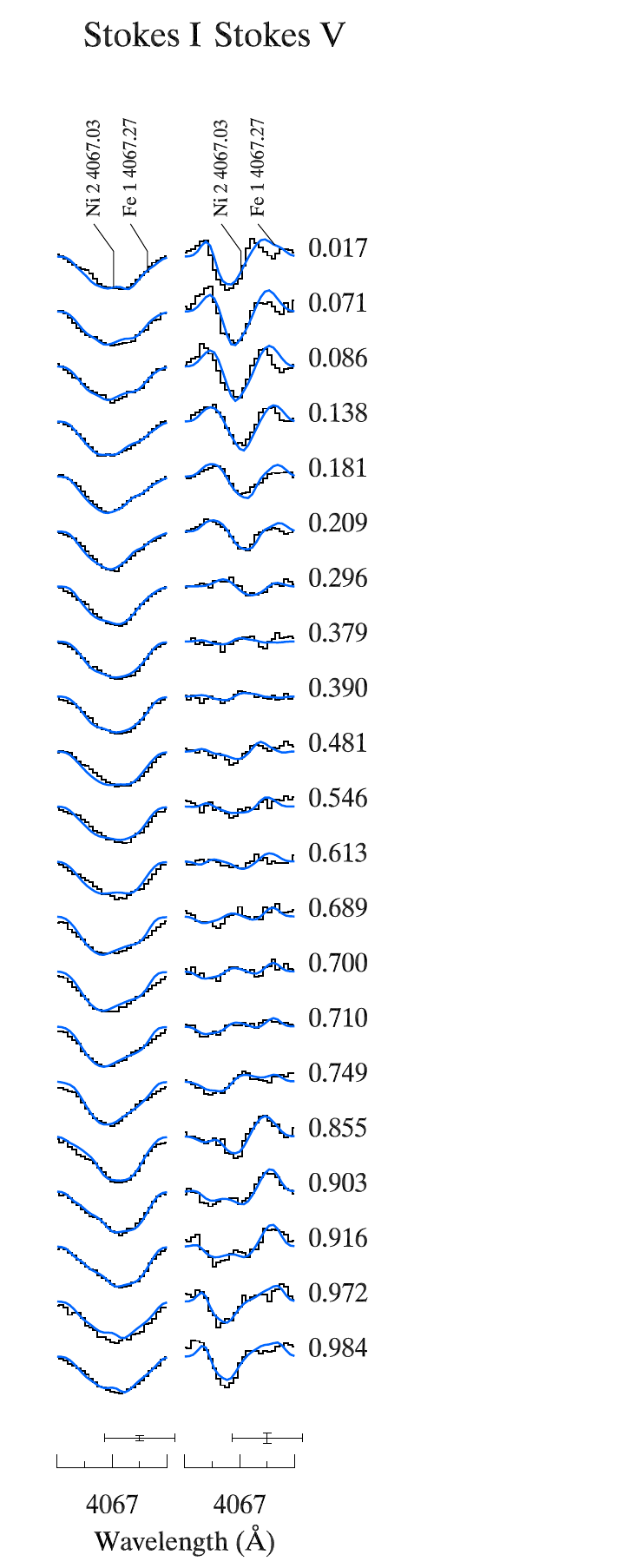}
   \caption{Same as Fig.~\ref{Fit-IV-Fld} for the observed (black histogram line) and synthetic (solid blue line) Stokes $IV$ spectra of Mg, Si, Ti and Ni lines used for abundance mapping of these elements.}
\label{Mg-Si-Ti-Ni-fit}
\end{center}
\end{figure*}

The titanium chemical map was reconstructed using the Stokes $IV$ profiles of the Ti~{\sc ii} $\lambda$ 4290 and 4563 lines. The fit between observations and the model can be seen in Fig.~\ref{Mg-Si-Ti-Ni-fit} and the resulting map is presented in Fig.~\ref{Maps-Abn}. The titanium abundance ranges from $-4.0$ dex to $-7.0$ dex  (the solar value is $-7.09$). The largest areas of Ti abundance enhancement are located close to the stellar equator, with distinct low abundance regions in the vicinity of the rotational poles. The areas of enhanced titanium are located where the field strength is somewhat weaker than the average, approximately 4 to 6 kG.  Some of the depleted abundance areas are located close to the regions of the strongest magnetic field. 

The chromium map was produced using the Stokes $IQUV$ profiles of the Cr~{\sc ii} $\lambda$ 4558, 4588 and 4824 lines. The comparison between observations and the model profiles is shown in Figs.~\ref{Fit-IV-Fld} and \ref{Fit-QU-Fld}. The corresponding map is displayed in Fig.~\ref{Maps-Abn}. The chromium abundance ranges from $-3.0$ dex to $-6.0$ dex  (solar value is $-6.40$).  Similar to what is found for titanium,  the high abundance regions are located in large structures. In this case, three distinct features can be identified, with two located at or close to the stellar equator and one (seen at phase 0.0) extending from the equator almost to the pole.  There are also two small depleted regions which are located similarly to the Ti depleted areas. Again, similar to Ti, it appears that areas of enhanced chromium are coincident with the regions where the field is weaker than the average. The small depleted abundance areas are located close to where the field is strongest. 

The iron map was obtained from the Stokes $IQUV$ profiles of the Fe~{\sc ii} $\lambda$ 4583, 4923, 5018 and 5169 lines. The observed and synthetic spectra are compared in Figs.~\ref{Fit-IV-Fld} and \ref{Fit-QU-Fld}, with the resulting map shown in Fig.~\ref{Maps-Abn}.  The iron abundance ranges from $-2.0$ dex to $-5.0$ dex  (the solar value is $-4.54$). Similar to what is found for titanium and chromium, the high abundance regions are located in large structures close to the stellar equator. Also similar to what is seen for titanium, there are smaller depleted regions which are located closer to the poles. Comparing the magnetic field structure and the iron abundance map, while a clear correlation is absent, it does appear that the large enhancement areas correlate with the relatively weak field regions while the small depleted abundance areas are located close to where the field is strongest. 

\subsection{Nickel and Neodymium} 

The nickel map was produced using the Stokes $IV$ profiles of the Ni~{\sc ii} $\lambda$ 4067 line.  The observed and best-fitting synthetic profiles are presented in Fig.~\ref{Mg-Si-Ti-Ni-fit} and the resulting map is shown in Fig. \ref{Maps-Abn}. The nickel abundance ranges from $-4.5$ dex to $-6.5$ dex  (the solar value is $-5.82$). The map appears to be quite complex. High abundances are localised to small structures,  with the biggest structure seen close to the stellar equator at phase 0.80.  For nickel there is no clear correlation between the location of abundance enhancements or depletion and the magnetic field structure.

The neodymium map was derived using the Stokes $IQUV$ spectra of Nd~{\sc iii} $\lambda$ 5050 and 5102.  The fit between observations and the model profiles can be seen in Figs.~\ref{Fit-IV-Fld} and \ref{Fit-QU-Fld}. The resulting map is shown in Fig. \ref{Maps-Abn}. The neodymium abundance ranges from $-5.5$ dex to $-8.0$ dex  (solar value is $-10.62$).  The abundance distribution of this element is fairly homogenous, with enhanced abundance area forming one spot just above the stellar equator at phase 0.0.  This area of Nd enhancement appears to coincide with the region of large negative radial magnetic field component.

\section{Conclusions and Discussion}
\label{sect:discuss}

In this study we presented the magnetic field maps and chemical abundance distributions for the Ap star HD\,32633. This is only the fourth Ap star for which the surface structure is derived by interpreting high-resolution spectropolarimetric data in all four Stokes parameters. Our magnetic maps of HD\,32633 reveal a largely axisymmetric, bipolar field topology. The poloidal $\ell=1$ harmonic mode is the dominant contributor, with nearly $75 \%$ of the energy in that mode; $9\%$ of the magnetic energy is contained in the poloidal quadrupolar $\ell=2$ mode. We also find $16 \%$ of the energy in toroidal components.  At the same time, we have shown that the Stokes parameter profiles of HD\,32633 cannot be fit by assuming a pure dipole or dipole + quadrupole magnetic field geometry. This indicates that the field is definitely more complex than expressed by such low-order multipolar parametrisation, despite a small contribution of $\ell\ge3$ modes in the harmonic power spectrum.

It is useful to compare our MDI results for HD\,32633 with the field geometries of other A and B stars studied using similar magnetic inversion methodologies. In doing so it must be kept in mind that MDI limited to Stokes $IV$ data loses some degree of field complexity compared to inversions based on the Stokes $IQUV$ spectra \citep{Kochukhov10,Rosen15}. Therefore, magnetic inversions with circular polarisation data provide a reliable information about the large scale field components but are less trustworthy when it comes to characterising the small-scale ($\ell\ge3$) magnetic structures superimposed on the global dipole-like field background.

With these caveats in mind, the late-B star CU\,Vir (HD\,124224) mapped using Stokes $IV$ \citep{Kochukhov14} was found to have 52--60$\%$ of the magnetic energy in the $\ell=1$ harmonic mode, and 23--30$\%$ in $\ell=2$ and $\ell=3$. This means that the global field topology of CU\,Vir is somewhat more complex but still comparable to that of HD\,32633.

A series of four Stokes parameter MDI studies of $\alpha^2$~CVn \citep{Kochukhov10,Silvester14a,Silvester14b} was performed using the version of {\sc Invers10} with the discrete field parameterisation. However, reconstructing the field topology with the input data of \citet{Silvester14b} and the same harmonic code as used in the present study, we find 73\% of the energy in $\ell=1$ and 10\% in $\ell=2$, which is almost identical to what was found for HD\,32633. On the other hand, the total contribution of the toroidal components is only $\approx$\,5 \% for $\alpha^2$~CVn compared to 16 \% for HD\,32633. Thus, we find that the degree of complexity of the global poloidal field structure of CU\,Vir, $\alpha^2$~CVn and HD\,32633 is similar. All three stars have spectral classes B9--A0 and masses in the narrow range of 3.0--3.1$M_\odot$ \citep{Kochukhov06}.

Cooler and less massive Ap stars tend to show simpler magnetic field topologies. The field in the extreme Ap star HD\,75049 \citep{Kochukhov15} was mapped using Stokes $IV$ and was found to be almost entirely poloidal (96\% of the magnetic energy in the poloidal harmonic components) and mostly dipolar (90\% of the energy in $\ell=1$ mode). The magnetic field of the roAp star HD\,24712 (DO~Eri) was investigated by \citet{Rusomarov15} using Stokes $IQUV$. They found 96\% of the energy in the $\ell=1$ mode with entirely poloidal structure. Thus, both of these stars have a simpler field configuration than found for HD\,32633, CU\,Vir and $\alpha^2$~CVn.

Examples of significantly more complex global magnetic field configurations are limited to substantially more massive B-type stars. For the helium-strong B2 star HD\,37776 (V901~Ori) mapped using Stokes $IV$ \citep{Kochukhov11} it was found that the largest contribution to the field topology was in the non-axisymmetric $\ell$\,=\,3--4 modes and that the field of that star contained a sizeable toroidal component. The young massive (B0.2V) star $\tau$~Sco (HD\,149438), studied with Stokes $IV$ \citep{Donati06}, also exhibits a complex field topology, with the $\ell=4$ mode being the strongest contributor to the magnetic field structure. Another Stokes $IV$ analysis of the B2 He-strong star HD\,184927 (V1671~Cyg) \citep{Yakunin15} revealed 38\% of the magnetic energy in the $\ell=1$ mode compared to 62\% in $\ell=2$, although higher order modes could not be meaningfully characterised for this star due to its a low inclination angle.

A few other A and B-type stars have been mapped with MDI, but without the benefit of characterising the field topology with spherical harmonics. The magnetic field geometry of 2.1$M_\odot$ Ap star 53\,Cam (HD\,65339) was studied by \citet{Kochukhov04} using lower quality MuSiCoS four Stokes parameter data. They found a dipolar-like field structure distorted by small-scale local spots, not unlike the field of $\alpha^2$~CVn. 

The B2 He-strong star $\sigma$~Ori~E (HD\,37479) was mapped with Stokes $IV$ data by \citet{Oksala15}. This star was found to have mostly dipolar magnetic field with a small non-axisymmetric quadrupolar contribution. Direct visual comparison of the derived magnetic field maps shows that the radial field of $\sigma$~Ori~E is less complex than that of HD\,32633.  Overall, the magnetic field geometry of $\sigma$~Ori~E is appears to be simpler than the field in HD\,37776, $\tau$~Sco and HD\,184927.

The different levels of the global magnetic field complexity of A and B stars studied with spherical harmonic MDI are summarised in Table~\ref{field-complexity}. The three columns in this table correspond to distinct mass ranges. So examining Table~\ref{field-complexity}, it is tempting to conclude, as already suggested by \citet{Rusomarov15}, that field complexity increased with stellar mass. However, this sample of magnetic stars is very small and inhomogeneous in the sense that only three out of eight objects were studied in all four Stokes parameters.

Theoretical modelling by \citet{Braithwaite09} showed that it is possible to have a stable magnetic field configuration for a predominantly poloidal field with some toroidal field components in the stellar interior, in essence a bipolar-like magnetic field structures as seen in HD 32633 and $\alpha^2$~CVn. At the same time it has also been shown by \citet{Braithwaite08} that more complex configurations, such as seen in  $\tau$~Sco, could result in a stable field. 

In addition to mapping the magnetic field of HD 32633 we also presented chemical abundance maps for 8 elements. Similar to what was found in the case of $\alpha^2$ CVn \citep{Silvester14b}, we find no clear correlation between the location of enhanced and depleted abundance areas and the magnetic field. What we do find however is that the enhanced areas of iron-peak elements and silicon seem to coincide with the areas where the field is weaker, which was also found for $\alpha^2$ CVn.

Theoretical diffusion modelling of rare earth elements and iron-peak elements by \citet{Michaud81} suggests that rare-earth elements should be concentrated where the magnetic field is horizontal and that iron should be enhanced where the field is vertical.  Similarly theoretical modelling of silicon by \citet{Alecian81} expects silicon to be enhanced at the magnetic equator. It has been suggested in the case of a magnetic field which was not purely dipolar,  that instead of belt-like enhancements seen at the magnetic equator,  only spotlike features should be observed at the equator with MDI \citep{Alecian10}.  None of these theoretical predictions are observed in our maps,  suggesting further work is required in the theoretical modelling of diffusion in the atmospheres of Ap stars.

\begin{table}
\begin{center}
\caption{Qualitative comparison of the magnetic field topology complexity of HD\,32633 with other stars studied with MDI.}
\begin{tabular}{ccc}
\hline
\hline
More Complexity  & Similar Complexity & Less Complexity \\
\hline
HD\,37776 & CU\,Vir & HD\,24712  \\
$\tau$~Sco & $\alpha^2$~CVn & HD\,75049 \\
HD\,184927 &  HD\,32633 &  \\
 \hline
\label{field-complexity}
\end{tabular}
\end{center}
\end{table}

\section*{Acknowledgments} 
OK is a Royal Swedish Academy of Sciences Research Fellow supported by grants from the Knut and Alice Wallenberg Foundation,  the Swedish Research Council and G\"{o}ran Gustafsson Foundation.
GAW acknowledges support from the Natural Science and Engineering Research Council of Canada in the form of a Discovery Grant.

\bibliographystyle{mnras}
\bibliography{astro_ref_v1}



\label{lastpage}

\end{document}